\documentclass[12pt,a4paper]{article}

\usepackage{ifthen} 
\usepackage{subfig}
\usepackage{lscape}
\usepackage{multirow}
\usepackage{placeins}
\newboolean{pdflatex}
\setboolean{pdflatex}{true} 

\newboolean{articletitles}
\setboolean{articletitles}{true} 

\newboolean{uprightparticles}
\setboolean{uprightparticles}{false} 

\FloatBarrier

\usepackage{ifpdf}
\ifpdf

\usepackage{hyperref}    
\usepackage{microtype}
\usepackage{lineno}  
\usepackage{xspace} 
\usepackage{caption} 

\usepackage{graphicx}  
\usepackage{color}
\usepackage{colortbl}
\graphicspath{{./figs/}} 

\usepackage{amsmath} 
\usepackage{amssymb}
\usepackage{amsfonts}
\usepackage{upgreek} 

\newcommand*\patchAmsMathEnvironmentForLineno[1]{%
\expandafter\let\csname old#1\expandafter\endcsname\csname #1\endcsname
\expandafter\let\csname oldend#1\expandafter\endcsname\csname
end#1\endcsname
 \renewenvironment{#1}%
   {\linenomath\csname old#1\endcsname}%
   {\csname oldend#1\endcsname\endlinenomath}%
}
\newcommand*\patchBothAmsMathEnvironmentsForLineno[1]{%
  \patchAmsMathEnvironmentForLineno{#1}%
  \patchAmsMathEnvironmentForLineno{#1*}%
}
\AtBeginDocument{%
\patchBothAmsMathEnvironmentsForLineno{equation}%
\patchBothAmsMathEnvironmentsForLineno{align}%
\patchBothAmsMathEnvironmentsForLineno{flalign}%
\patchBothAmsMathEnvironmentsForLineno{alignat}%
\patchBothAmsMathEnvironmentsForLineno{gather}%
\patchBothAmsMathEnvironmentsForLineno{multline}%
\patchBothAmsMathEnvironmentsForLineno{eqnarray}%
}

\def\lhcb {\mbox{LHCb}\xspace}
\def\atlas  {\mbox{ATLAS}\xspace}
\def\cms    {\mbox{CMS}\xspace}
\def\alice  {\mbox{ALICE}\xspace}

\def\MagUp {\mbox{\em Mag\kern -0.05em Up}\xspace}

\ifthenelse{\boolean{uprightparticles}}%
{
 
 \def\Pgamma      {\ensuremath{\upgamma}\xspace}

 \def\Ppi         {\ensuremath{\uppi}\xspace}

 \def\Ptau        {\ensuremath{\uptau}\xspace}

 \def\PDelta      {\ensuremath{\Delta}\xspace}                 
 \def\PXi      {\ensuremath{\Xi}\xspace}                 
 \def\PLambda      {\ensuremath{\Lambda}\xspace}                 
 \def\PSigma      {\ensuremath{\Sigma}\xspace}                 
 \def\POmega      {\ensuremath{\Omega}\xspace}                 
 \def\PUpsilon      {\ensuremath{\Upsilon}\xspace}                 
 
 \def\PB      {\ensuremath{\mathrm{B}}\xspace}                 
                  
 \def\PD      {\ensuremath{\mathrm{D}}\xspace}

 \def\PK      {\ensuremath{\mathrm{K}}\xspace}

 \def\PW      {\ensuremath{\mathrm{W}}\xspace}

 \def\PZ      {\ensuremath{\mathrm{Z}}\xspace}                 
                  
 \def\Pb      {\ensuremath{\mathrm{b}}\xspace}

 \def\Pi      {\ensuremath{\mathrm{i}}\xspace}

}
{
 
 \def\Pgamma      {\ensuremath{\gamma}\xspace}

 \def\Ppi         {\ensuremath{\pi}\xspace}

 \def\Ptau        {\ensuremath{\tau}\xspace}

 \mathchardef\PDelta="7101
 \mathchardef\PXi="7104
 \mathchardef\PLambda="7103
 \mathchardef\PSigma="7106
 \mathchardef\POmega="710A
 \mathchardef\PUpsilon="7107
                  
 \def\PB      {\ensuremath{B}\xspace}                 
                  
 \def\PD      {\ensuremath{D}\xspace}

 \def\PK      {\ensuremath{K}\xspace}

 \def\PW      {\ensuremath{W}\xspace}

 \def\PZ      {\ensuremath{Z}\xspace}                 
                  
 \def\Pb      {\ensuremath{b}\xspace}

 \def\Pi      {\ensuremath{i}\xspace}

}

\makeatletter
\ifcase \@ptsize \relax
  \newcommand{\miniscule}{\@setfontsize\miniscule{4}{5}}
\or
  \newcommand{\miniscule}{\@setfontsize\miniscule{5}{6}}
\or
  \newcommand{\miniscule}{\@setfontsize\miniscule{5}{6}}
\fi
\makeatother

\DeclareRobustCommand{\optbar}[1]{\shortstack{{\miniscule (\rule[.5ex]{1.25em}{.18mm})}
  \\ [-.7ex] $#1$}}

\def\W      {{\ensuremath{\PW}}\xspace}

\def\Z      {{\ensuremath{\PZ}}\xspace}

\def\bquark    {{\ensuremath{\Pb}}\xspace}
\def\bquarkbar {{\ensuremath{\overline \bquark}}\xspace}
\def\bbbar     {{\ensuremath{\bquark\bquarkbar}}\xspace}

\def\pion   {{\ensuremath{\Ppi}}\xspace}
\def\piz    {{\ensuremath{\pion^0}}\xspace}

  \def\Kbar    {{\kern 0.2em\overline{\kern -0.2em \PK}{}}\xspace}

\def\KorKbar    {\kern 0.18em\optbar{\kern -0.18em K}{}\xspace}

  \def\Dbar    {{\kern 0.2em\overline{\kern -0.2em \PD}{}}\xspace}

\def\DorDbar    {\kern 0.18em\optbar{\kern -0.18em D}{}\xspace}

\def\Bbar    {{\ensuremath{\kern 0.18em\overline{\kern -0.18em \PB}{}}}\xspace}

\def\BorBbar    {\kern 0.18em\optbar{\kern -0.18em B}{}\xspace}

  \def\Y#1S{\ensuremath{\PUpsilon{(#1S)}}\xspace}

\def\Lbar        {{\ensuremath{\kern 0.1em\overline{\kern -0.1em\PLambda}}}\xspace}
\def\LorLbar    {\kern 0.18em\optbar{\kern -0.18em \PLambda}{}\xspace}

\def\ra                 {\ensuremath{\rightarrow}\xspace}

\def\order   {{\ensuremath{\mathcal{O}}}\xspace}

\def\AT#1     {\ensuremath{A_{\mathrm{T}}^{#1}}\xspace}           

\def\C#1      {\ensuremath{\mathcal{C}_{#1}}\xspace}                       
\def\Cp#1     {\ensuremath{\mathcal{C}_{#1}^{'}}\xspace}                    
\def\Ceff#1   {\ensuremath{\mathcal{C}_{#1}^{\mathrm{(eff)}}}\xspace}        
\def\Cpeff#1  {\ensuremath{\mathcal{C}_{#1}^{'\mathrm{(eff)}}}\xspace}       
\def\Ope#1    {\ensuremath{\mathcal{O}_{#1}}\xspace}                       
\def\Opep#1   {\ensuremath{\mathcal{O}_{#1}^{'}}\xspace}                    

\newcommand{\tev}{\ensuremath{\mathrm{\,Te\kern -0.1em V}}\xspace}
\newcommand{\gev}{\ensuremath{\mathrm{\,Ge\kern -0.1em V}}\xspace}
\newcommand{\mev}{\ensuremath{\mathrm{\,Me\kern -0.1em V}}\xspace}
\newcommand{\kev}{\ensuremath{\mathrm{\,ke\kern -0.1em V}}\xspace}
\newcommand{\ev}{\ensuremath{\mathrm{\,e\kern -0.1em V}}\xspace}
\newcommand{\gevc}{\ensuremath{{\mathrm{\,Ge\kern -0.1em V\!/}c}}\xspace}
\newcommand{\mevc}{\ensuremath{{\mathrm{\,Me\kern -0.1em V\!/}c}}\xspace}
\newcommand{\gevcc}{\ensuremath{{\mathrm{\,Ge\kern -0.1em V\!/}c^2}}\xspace}
\newcommand{\gevgevcccc}{\ensuremath{{\mathrm{\,Ge\kern -0.1em V^2\!/}c^4}}\xspace}
\newcommand{\mevcc}{\ensuremath{{\mathrm{\,Me\kern -0.1em V\!/}c^2}}\xspace}

\def\mum  {\ensuremath{{\,\upmu\rm m}}\xspace}

\def\sec  {\ensuremath{\rm {\,s}}\xspace}

\def\order{{\ensuremath{\cal O}}\xspace}

\def\gsim{{~\raise.15em\hbox{$>$}\kern-.85em
          \lower.35em\hbox{$\sim$}~}\xspace}
\def\lsim{{~\raise.15em\hbox{$<$}\kern-.85em
          \lower.35em\hbox{$\sim$}~}\xspace}

\def\pt         {\mbox{$p_{\rm T}$}\xspace}

\def\fastjet    {\mbox{\textsc{FastJet}}\xspace}
\def\antikt     {\ensuremath{\rm anti\mbox{-}k_T}\xspace} 
\def\kt         {\ensuremath{\rm k_T}\xspace}

\def\tell1  {TELL1\xspace}
\def\ukl1   {UKL1\xspace}

\usepackage{cite} 
\usepackage{mciteplus}

\begin{document}

\renewcommand{\thefootnote}{\fnsymbol{footnote}}
\setcounter{footnote}{1}

\begin{titlepage}

\vspace*{-1.5cm}

\hspace*{-0.5cm}
\begin{tabular*}{\linewidth}{lc@{\extracolsep{\fill}}r}
\ifthenelse{\boolean{pdflatex}}
{\vspace*{-2.7cm}\mbox{\!\!\!} & &}%
{\vspace*{-1.2cm}\mbox{\!\!\!} & &}
 \\
 & & New Trends of High Energy Physics \\  
\hline
\end{tabular*}

\vspace*{4.0cm}

{\bf\boldmath\huge
\begin{center}
  Experimental aspects of jet physics at LHC
\end{center}
}

\vspace*{2.0cm}

\begin{center}
Murilo Rangel
\bigskip\\
{\it\footnotesize
Universidade Federal do Rio de Janeiro
}
\vfill
\end{center}

\vspace{\fill}

\begin{abstract}
  \noindent
  Jet physics provides a powerful tool to investigate interaction properties of quarks and gluons. These studies have been possible at an energy never investigated before at LHC. In this proceedings we review the main characteristics of experimental methods to measure jets in proton–proton collisions at center-of-mass energies of 7 and 8 TeV. Novel methods are expected to play an important role for searching new physics at center-of-mass energy of 13 TeV.
\end{abstract}

\vspace*{2.0cm}
\vspace{\fill}

\end{titlepage}

\pagestyle{empty}  





\renewcommand{\thefootnote}{\arabic{footnote}}
\setcounter{footnote}{0}



\pagestyle{plain} 
\setcounter{page}{1}
\pagenumbering{arabic}


\section{Introduction}
\label{sec:introduction}

Jets are the signatures of quarks and gluons produced in high-energy collisions such as the proton-proton interactions at the Large Hadron Collider (LHC). The understanding of the jet properties are key ingredients of several physics measurements and for New Physics searches.  The study of jets have been used to test perturbative QCD (pQCD), to probe proton structure and to search for New Physics. 

This paper is a summary of two lectures given in the school {\it{New Trends in High-Energy Physics and QCD}}. The slides can be found in this \href{http://indico.cern.ch/event/346738/contribution/52/attachments/684118/939687/14_JetPhysics.pdf}{link}. Deliberately no figure is used here and the focus will be on the main ingredients of performing jet phyics at LHC and novel techniques introduced in RunI.

This paper is organized as follows. The general aspects of experimental inputs for jet reconstruction are described in Sec.~\ref{sec:inputs}. Section~\ref{sec:reconstruction} introduces the jet reconstruction. Jet energy measurement is discussed in section~\ref{sec:jec}. Novel jet physics methods studied at LHC RunI are presented in section~\ref{sec:novel}. A summary is found in section~\ref{sec:summary}.

\section{Particles from detector}
\label{sec:inputs}

The first step of any jet analysis is the particles to be used in the jet reconstruction algorithm. Two main approaches are used by LHC experiments: calorimeter measurements and particle flow candidates. Calorimeter (CALO) jets are reconstructed from energy deposits in the calorimeter clusters while particle flow (PF) jets are reconstructed from from particles identified from different sub-detectors. The main differences from these approaches will be discussed below.

The four main detectors at LHC are capable of measuring energy and hits of emerging particles of the high-energy collision with proper time bigger than $10^{-8}\sec$. Detailed information of the \alice, \atlas, \cms and \lhcb detectors can be found elsewhere~\cite{Alice,Atlas,CMS,Alves:2008zz}. From these measurements, particle momentum and identification can be inferred with good precision. With a certain level of generalisation, we can say five types of particles are identified: photons, electrons, muons, charged hadrons and neutral hadrons. In the PF algorithm, particle identification is the main feature to provide information of the jet characteristics. 

Muons are the easiest particles to identify in high-energy physics detectors. Since they are heavy and do not interact via quantum chromodynamics, they traverse all the sub-detectors and produce hits in the so-called muon stations. 

Photons are identified as clustes in the eletromagnetic calorimeter with void of hits in the track sub-detectors. The electrons produce similar showers in the eletromagnetic calorimeter, but a reconstructed track is used as main discrimination from photon showers. To reduce interaction of hadrons in the eletromagnetic calorimeter, low density with high-Z material is used as absorber. In more complex algorithms, photon convertion in the track sub-detectors is also considered.

Neutral and charged hadrons are reconstructed in the hadronic calorimeter where low nuclear interaction length is used as absorber. When a track extrapolated to the hadronic calorimeter matches with a cluster, the track is taken as charged hadron. Energy in the hadron calorimeter that is matched with a track is not used since the track momentum resolution is better than calorimeter energy resolution. In more detailed algorithms, V0's decays are identified from displaced vertices and \piz\ra\Pgamma\Pgamma are selected using the eletromagnetic calorimeter.

Jets reconstructed from PF-particles have usually better energy resolution and smaller calibration factors, but they have properties harder to model due to the heterogeneity of the inputs. CALO-jets have in general well understood modeling and their energy resolution and scale are improved with indirect information from other sub-detectors, e.g., calorimeter is calibrated using tracks.

Reconstruction of jets using stable particles produced by a given event generator is used to calibrate jets reconstructed from detector inputs. Jet physics observables can only have a meaning if detector-level jets are calibrated. Discussion on jet calibration is presented in section~\ref{sec:jec}.

\section{Jet reconstruction}
\label{sec:reconstruction}

LHC experiments use in general reconstruction algorithms implemented in the \fastjet package~\cite{FastJet}. The most widely used algorithms are the \antikt~\cite{antikt}, \kt~\cite{kt1,kt2,kt3} and the Cambridge-Aachen (C/A) algorithm~\cite{ca1,ca2}. The \kt and C/A algorithms provides spatial and kinematic information about the substructure of jets, since they carry the clustering history. The \antikt algorithm define jets using successive recombinations providing almost no information about the \pt ordering of the shower. Therefore, the \kt and C/A algorithms are usually used for jet substructure studies, while \antikt is used to study single-parton jet physics.

The jet algorithms do not reject jets originating from detector noise, pile-up particles, high-\pt leptons, hadronic \Ptau decays and cosmic rays. Criteria to reject fake and noise jets are used called jet identification (JetID). Background jets rate are reduced to $\mathcal{O}$(1\,\%) by using jet properties, e.g., charged energy fraction (see studies performed by CMS as example~\cite{CMS-PAS-JME-09-008}).

Many interesting physics processes at LHC have bottom or charm quark production, e.g., Higgs and top quark decays. Therefore, identifying jets originating from bottom or charm quarks becomes a very important task. Fragmentation of bottom and charm quarks produces hadrons with relatively large masses, long lifetimes and high-\pt charged particles. Besides good track momentum resolution, the detector needs to have precise reconstruction of the secondary vertex of \order(30\mum). These aspects are explored in various algorithms that have about 50\,\% tagging efficiency for 1\,\% fake rate. ALICE, ATLAS, CMS and LHCb have documented their studies of bottom and charm quark jet identification in ref.~\cite{Atlasbtag,CMSbtag,LHCb-PAPER-2015-016,Alicebtag}.

Discrimination between light quark- and gluon-initiated jets can be very helpful when searching for new processes with many light quarks \cite{ATLASqg,CMSqg}. Since the color factor of the gluon is 3 and for light quarks ($u,d$ and $s$) is 4/3, it is expected that the number of particles produced in gluon-initiated jets is $9/4$ times than in light quark initiated jets. The width of gluon-initiated jets are also expected to be larger than light quark-initiated jets. In general, these jet characteristics are explored in multivariate discriminant that provide 60\,\% efficiency for a fake rate of 30\,\%.

\section{Jet energy measurement}
\label{sec:jec}

The jet energy clustered from the reconstructed particles differs from the corresponding true jet energy clustered from the stable particles before interacting with the detector. Huge part of effort of understanding jets is the jet energy calibration, i.e., correct the detector-level jet to the particle-level jet. In other words, the jet energy measurement becomes independent of the detector. Since jets are a collection of particles or calorimeter clusters, jet energy calibration is a very difficult task (see ref.~\cite{CMSjec,Atlasjec}).

In general, the calibration is factorized in different factors: offset, relative and absolute. In some cases, the jet direction is also corrected for which can be up $\mathcal{O}$(1\,\%) correction. Dedicated jet energy calibration (JEC) can also be derived for different parton where the jet originates from.

The offset correction is usually derived in data-driven studies and treated as a linear correction with the number of vertices reconstructed in the event. This assumption works well for CALO-jets, but PF-jets need to use more sophisticated methods, e.g., jet area~\cite{jetarea}. The main goal of this factor is to subtract the energy not associated with the high-\pt scattering. Most of the energy excess originates from pile-up or out-of-time events. This factor becomes extreme important at high luminosity conditions.

The relative factor explores the best region of the detector for jet reconstruction. The goal is to calibrate regions of the detector with poor resolution with respect to the best understood region. One of the main advantages of this procedure is that no simulation is needed. The most used technique is the dijet \pt-balance. Due the high cross-section and the two jet well balanced in \pt, the dijet production provides an unique sample to measure the relative response of jet energy measurement.

The absolute correction factor aims the calibration of the jet energy to the true jet, i.e., the energy that would have been measured by a perfect detector. This factor is in general derived in simulation and residual corrections are estimated with data. Two event samples are usually used to compare data with simulation: \Pgamma+jets and \Z+jets. The method is based on the correlation between the jet \pt and the \Pgamma or \Z \pt. While the \Pgamma+jets sample provides much more statistics, the size of backgrounds is also larger. Besides the background contamination and event sample size differences, \Pgamma and \Z have also different energy resolutions. Compromise of background, size sample and resolution of the reference object needs to be studied.

Dedicated calibration for jets originating from bottom quarks can also be derived using \Z+b-jets, top quark or \Z\ra\bquark\bquarkbar decays (for example, see ref.~\cite{CMSbjec}). Differences between different parton-initiated jets can be up to 2\,\%. Most of experiments include this effect in the JEC error.

After calibrating the jet energy, precise knowledge of its resolution is crucial for jet physics. In differential cross-section measurements, the jet energy measurement needs to be unfolded to the true jet energy, and the main impact of the unfolding procedure is the jet energy resolution (JER). Other analyses need good understanding of the differences between data and simulation, e.g., Higgs decaying to \bbbar. JER can studied with dijet, \Pgamma+jets and \Z+jets samples (see refs.~\cite{CMSjec,Atlasjer}). In the case of dijet events, the impact of the final state radiation in the dijet \pt asymmetry must be done by using different \pt thresholds for the third jet. The extrapolation to zero \pt threshold provides the true jet energy resolution.

W/Z and top quark masses are known with relatively high precision, therefore, hadronic decays of these particles can be used to certify jet energy measurements.

\section{Novel methods using jets}
\label{sec:novel}

It is crucial to develop novel methods to improve the impact of experimental results using the available data. LHC RunII data will provide great opportunity to apply novel jet methods. Many ideas are already successfully tested with RunI data and we shall summarize them below.

Various theoretical ideas has been developed by using large are jets or fat jets~\cite{fatjet1,fatjet2,fatjet3}. When an unstable particle is produced at \pt greater than twice its mass, their decay products are produced collimated with respect to the beam axis. The LHC data probes kinematic regimes with the production of SM particles with significant Lorentz boosts, or even new massive particles that decay to highly boosted SM particles. For example, when sufficiently boosted, the decay products of \W bosons, top quarks, and Higgs bosons can become collimated to the point that standard jet reconstruction techniques fail. In other words, when the separation of the quarks in these boosted topologies is smaller than the radius parameter of the jet reconstruction, individually resolved jets can not be identified. At RunI, many studies provided promising results for the use of these techniques at RunII.

Boosted \W production was studied by CMS~\cite{CMSWboost} using the C/A algorithm with distance parameter $R=0.8$ and jets with \pt greater than 200\gev. The \W boson is selected to be a decay product of a top quark and another leptonically decaying \W boson in the event is selected. The \W-jet is then submitted to various tests that ensure it has substructure. The typical efficiency of this method to tag a boosted \W is 65\,\% with a background rejection of 96\,\%.

ATLAS collaboration also published studies with boosted top quarks, \W and \Z bosons~\cite{ATLASboost}. In general, the C/A algorithm provides better framework to tag boosted objects, but \kt and \antikt are also used to in specific cases. The main challenge is to suppress parton-initiated jets while keeping the signal mass peak unaffected. Techniques as the grooming algorithms show great performance on this task.

Searches for new physics using RunI data and boosted objects were performed by both CMS and ATLAS experiments~\cite{ATLASboostsearch1,CMSboostsearch1}. Heavy particles ranging with masses up to 2000\gev can be probed in these analyses. Jet substructure simulation studies at very high luminosity collisions show promising prospects for the future results with RunII data.

ALICE experiment measured the ratio of the inclusive differential jet cross sections for $R = 0.2$ and $R = 0.4$ \cite{Alicejet1}. This ratio allows a more stringent comparison of data and calculations than the individual inclusive cross sections \cite{soyezsimple}, since many systematic uncertainties are common or highly correlated. The pQCD calculation considers the ratio directly, rather than each distribution separately, making the calculated ratio effectively one perturbative order higher than the individual cross sections \cite{soyezsimple}. This is nice example novel methos in jet physics can help to test precisely the SM.

\section{Summary}
\label{sec:summary}

This letter highlighted the main ingredients to build jets and the novel ideas explored in RunI. Jet physics played a major role at LHC RunI. Precise tests of the SM and possible discoveries with RunII data will certainly depend on the jet tools.

\section*{Acknowledgements}

\noindent 
The author would like to thank the organisers and in particular Christophe Royon for the invitation. The author acknowledges support from the brazilian agencies CAPES, CNPq and FAPERJ.

\addcontentsline{toc}{section}{References}
\bibliographystyle{LHCb}
\bibliography{main,LHCb-DP,LHCb-PAPER,LHCb-CONF}

\end{document}
\else
\fi